\documentclass{aa}

\usepackage{graphicx}
\usepackage{txfonts}
\usepackage{aalongtable,lscape}
\begin{document}
   \title{AGN counts at 15$\mu$m}

   \subtitle{XMM observations of the ELAIS-S1-5 sample}

   \author{F. La Franca,
          \inst{1}
          S. Puccetti\inst{2}
\fnmsep\thanks{INAF personell resident at ASDC},
N. Sacchi\inst{1},
C. Feruglio\inst{3}, 
F. Fiore\inst{3}, 
C. Gruppioni\inst{4},
A. Lamastra\inst{1}, 
I. Matute\inst{5},
G. Melini\inst{1},
F. Pozzi\inst{4,6}
     }

   \offprints{Fabio La Franca}

   \institute{Dipartimento di Fisica, Universit\`a Roma Tre,
              Via della Vasca Navale 84, I-00146 Roma, Italy\\
              \email{lafranca@fis.uniroma3.it}
         \and
ASI Science Data Center, via Galileo Galilei, I-00044, Frascati, Italy
\and
INAF-Osservatorio Astronomico di Roma, via Frascati 33, I-00040 Monteporzio 
Catone, Italy.
\and
INAF, Osservatorio Astronomico di Bologna, Via Ranzani 1, I-40127 Bologna, Italy
\and
INAF, Osservatorio Astrofisico di Arcetri, Largo E. Fermi 5,
I-50125, Firenze, Italy
\and
Dipartimento di Astronomia, Universit\`a di Bologna, 
         Via Ranzani 1, I-40127 Bologna, Italy
             }
   \date{June 2007; A\&A in press}

  \abstract
  {The counts of galaxies and AGN in the mid infra-red (MIR) bands are
    important instruments for studying their cosmological evolution. 
However, the classic spectral line
    ratios techniques can become misleading when trying to properly
    separate AGN from starbursts or even from apparently normal galaxies.}
  {We use X-ray band observations to discriminate AGN activity in
    previously classified MIR-selected starburst galaxies and to derive
    updated AGN1 and (Compton thin) AGN2 counts at 15 $\mu$m.}
  {XMM observations of the ELAIS-S1 15$\mu$m sample down to flux
    limits $\sim$ 2$\times$10$^{-15}$ erg cm$^{-2}$ s$^{-1}$ (2-10 keV
    band) were used. We classified as AGN all those MIR sources with a
    unabsorbed 2-10 keV X-ray luminosity higher that $\sim$10$^{42}$ erg
    s$^{-1}$.}
  { We find that at least about 13$\pm 6$\% of the previously
    classified starburst galaxies harbor an AGN. According to these
    figures, we provide an updated estimate of the counts of AGN1 and
    (Compton thin) AGN2 at 15 $\mu$m. It turns out that at least 24\% of the
extragalactic sources brighter than 0.6 my at 15$\mu$m are AGN ($\sim$13\%
contribution to the extragalactic background produced at fluxes brighter than 0.6 mJy). }
{}



\authorrunning{F. La Franca et al.}
   \maketitle
%

\section{Introduction}

To measure the evolution of the density of active galactic nuclei
(AGN) and galaxies at various wavelengths is one of the major goals in
astrophysics, and extragalactic counts in the mid-infrared (MIR) are
important instruments for this. MIR surveys, carried out with the
ISOCAM instrument on board the {\it Infrared Space Observatory} ({\it
  ISO}), indicates that the MIR galaxies have evolved significantly
faster than deduced from optical surveys (e.g.  Elbaz et al. 1999;
Gruppioni et al. 2002; Metcalfe et al. 2003; Rodighiero et al. 2004;
Pozzi et al. 2004). These results are now confirmed and extended by
the data coming from the {\it Spitzer Space Telescope} (e.g. Gruppioni
et al. 2005; Le Floc'h et al. 2005; P\'erez-Gonz\'alez et al.  2005;
Babbedge et al. 2006; Caputi et al. 2007). On the contrary, the
evolution of AGN at MIR wavelengths are similar to what has been
measured in the optical and X-ray bands for type 1 AGN (AGN1) and even
slower for type 2 AGN (Matute et al. 2002, 2006).

Proper analysis of the evolution of the sources emitting in the MIR is
based on a reliable classification that allows a separation of AGN
(more precisely type 2 AGN; AGN2) from the starburst
galaxies. Unfortunately there is growing evidence that the
classification of AGN based on their optical spectra alone provides an
incomplete and sometimes deceiving description of their true nature.
Indeed, some starbursts, or even normal, galaxies show signatures of a
hidden AGN at non-optical wavelengths (see e.g. Moorwood et al.  1996;
Marconi et al. 2000; Severgnini et al. 2003), while a fraction of
X-ray selected AGN show no signatures of AGN activity in their optical
spectra: the X-ray bright optically normal galaxies (XBONG, see
e.g. Fiore et al. 2000; Comastri et al.  2002; La Franca et al.  2005
and references therein).  Combining {\it Chandra} and archival X-ray
data, Maiolino et al.  (2003) have found that the local total density
of AGN may be a factor $\sim$2 higher than estimated from optical
spectroscopic surveys.  Up to now the exact proportion of
AGN-dominated MIR galaxies is actually not known.  This separation is
also complicated by the existence of mixed systems, where both star
formation and AGN activity significantly contribute to the mid-IR
emission (e.g. Fritz et al. 2006; Pozzi et al. 2007).

Mid-IR studies based on {\it IRAS} and {\it ISO} observations have
indirectly estimated a contribution of AGN to the cosmic infrared
background (CIRB) that is not larger than 5--10$\%$ (e.g. Franceschini
et al.  2001; Xu et al 2001, 2003). This is the maximum room left in
their models by the strongly evolving starburst population. The total
contribution was, in any case, uncertain since mid-IR selected AGN2
and starburst galaxies were treated as a single population.  A better
estimate of the AGN contribution comes from the X--ray band (0.5--10
keV), which offers a better wavelength regime for selecting and
identifying obscured sources (unless the sources are Compton-thick,
$N_H \! > \! 10^{24}-10^{25}$\,cm$^{-2}$).  A cross correlation of
X--ray and IR {\it ISO} sources detected in deep {\it Chandra} and
XMM-Newton observations allowed Fadda et al. (2002) to estimate that
the maximum fraction of the CIRB produced by AGN is 17$\pm$6 \%.  This
result is confirmed by Silva et al. (2004), who used the X-ray AGN
luminosity function (LF) and semi-empirical SEDs (linking the X--ray
to the Infrared) to derive that the contribution of AGN and their {\it
host} galaxy to the CIRB is 10-20\% in the 1-20 $\mu$m range.  These
results recently have been confirmed and extended by the analysis of
joint {\it Spitzer} and X-ray data, where a contribution of 3-11\% is
derived (see e.g. Treister et al. 2006; Barmby et al. 2006; Brand et
al. 2006).

On the other hand, it has been proposed that heavily X-ray obscured
sources could be selected in the mid-infrared band where the dust
heated by the AGN produces a spectral energy distribution (SED) that
peaks at shorter wavelengths than the SED produced by dust heated by
starbursts. Recently {\it Spitzer} data have been used to find highly
obscured AGNs, both trough mid-infrared color selection (e.g. Stern et
al. 2005; Polletta et al. 2006; Alonso-Herrero et al. 2006; Lacy et
al. 2007) and joint radio-infrared selection (e.g. Donley et al.
2005; Weedman et al. 2006; Martinez-Sansigre et al.  2005, 2006).
However, recent studies are showing that the situation is not simple.
The selection of AGN with mid-IR colors is showing up as not
simultaneously complete and reliable; and, moreover, when X-ray data
are available, the AGN mid-IR SED do not directly correspond to the
X-ray spectral type (absorbed/un-absorbed; e.g. Barmby et al.  2006;
Donley et al. 2007).

In this framework it is interesting to work out more reliable AGN1 and
AGN2 counts in the MIR using the 2---10 keV X-ray luminosity as a
solid indication of nuclear activity.  In this work we use XMM-Newton
observations of the S1 area of the {\it European Large Area ISO
  Survey} (ELAIS hereafter; Oliver et al.  2000; Rowan-Robinson et al.
2004) to revise the optical spectroscopic classification of
the 15 $\mu$m sample presented by La Franca et al. (2004) and then
to derive new, more reliable extragalactic counts of AGN1 and not heavily
absorbed (logN$_H$$<$24 cm$^{-2}$, not Compton thick) AGN2 in the
0.5--10 mJy regime, where no other ``class-separated'' data exist at
the moment.

The MIR and X-ray data are presented in \S 2. In \S 3, we describe the
X-ray properties of the MIR sources, while the revised AGN 15 $\mu$m
counts are derived in \S 4. Section 5 contains our comments and
conclusions.  Unless otherwise stated, all quoted errors are at the
68\% confidence level.  We assume H$_0$= 70 Km s$^{-1}$ Mpc$^{-1}$,
$\Omega_m$=0.3, and $\Omega_\Lambda$=0.7.

\section{The data}
\subsection{The 15 $\mu$m sample}

The mid-IR selected sources used in this work have been extracted from
the 15$\, \mu$m sample selected in the central region 5 of the field
S1 of the ELAIS. The 15$\, \mu$m catalogue in the ELAIS S1 field has
been released by Lari et al. (2001). It covers an area of $\sim$4
deg$^2$ centered at $\alpha(2000)=00^h 34^m 44.4^s$,
$\delta(2000)=-43^o 28' 12''$ and includes 462 mid-IR sources down to
a flux limit of 0.5 mJy. Mid-IR source counts based on this catalogue
have been presented and discussed by Gruppioni et al. (2002). La
Franca et al. (2004) presents the optical and spectroscopic
identifications and classification of a more reliable subsample of 406
sources from the Lari et al. (2001) catalogue. About 80$\%$ of these
sources have been optically identified on CCD exposures down to {\it
  R}$\sim$23. The spectral classification has been obtained for 90$\%$
of the optically identified sample. As discussed by La Franca et
al. (2004), due to a different mid-IR flux limit coverage of the
ELAIS-S1 field, the total area has been divided into two regions: the
central and deepest region of S1 (S1-5) reaching mid-IR fluxes
($S_{15}$ hereafter) of 0.5 mJy, and the outer region (S1-rest) with a
0.9 mJy flux limit. The S1-5 area is spectroscopically complete at the
97$\%$ level down to $R$=21.6, while S1-rest completeness reaches the
98$\%$ level down to $R$=20.5.  In total, 116 sources (29$\%$ of the
total mid-IR sample) do not have a spectroscopic identification due to
incompleteness of the follow-up or to the lack of optical counterpart
brighter than $R$=23. A detailed description of the optical
identification, spectroscopic classification, size, and completeness
function of the different areas used in the ELAIS-S1 sample are
presented and discussed by La Franca et al. (2004).

Classification for AGN-dominated sources in the ELAIS fields was based
on their optical spectral signatures.  Sources showing broad emission
line profiles (rest frame FWHM$>$1200 $km s^{-1}$) were classified as
type-1.  Type-2 sources were selected following classic diagnostic
diagrams (e.g. Tresse et al. 1996; Osterbrock 1989; Veilleux \&
Osterbrock 1987) that included one or more of the following line
ratios: [NII]/H$\alpha$, SII/(H$\alpha$+[NII]), OI/H$\alpha$,
[OIII]/H$\beta$, and [OII]/H$\beta$ when available, depending on the
redshift of the source ({\it e.g.}  log([OIII]/H$\beta$)$>$0.5 and
log([NII]/H$\alpha$)$>$-0.4).

Few objects have been re-observed in the framework of the follow-up
campaign of the {\it Spitzer} Wide-Area Infrared Extragalactic Survey
(SWIRE; see e.g. Lonsdale et al. 2003), which includes the ELAIS-S1
field. The ELAIS-S1 field has been observed in the B and V bands down
to 25 Vega mag. and in the R band down to 24.5 mag (Berta et al.
2006). A spectroscopic follow-up campaign was carried out in the
period 2004-2006 at the 3.6m and VLT ESO telescopes (La Franca et al.
in prep.). In comparison with La Franca et al. (2004), two more
15$\mu$m sources were spectroscopically identified in the ELAIS-S1-5
field: C15\_J003317-431706 is a R=24.3 galaxy showing [OII] emission
at redshift 0.689, while C15\_J003447-432447 is an AGN2 at redshift
1.076. Moreover, three sources with previously poor quality spectra
changed their spectroscopic classification: C15\_J003330-431553, which
was wrongly classified as a starburst galaxy at z=0.473, showed broad
CIII and MgII emission at z=2.170 typical of AGN1 activity;
C15\_J003603-433155, which was classified as AGN2, showed a broad MgII
emission typical of AGN1 activity; and C15\_J003622-432826, which was
classified as a starburst galaxy, showed a clear [OIII]/H$\beta$ ratio
typical of AGN2 activity.  In summary, the MIR sample used in this
work contains 72 bona fide extragalactic sources, of which 67 sources
have been optically identified down to R$\sim$25. Fifty-seven of them
have been spectroscopically identified: six are AGN1, three AGN2,
while almost all of the remaining 48 sources show the emission line
spectra typical of starburst activity. Most of the remaining 15
sources without spectroscopic identification (and for five sources
even optical identification) are so faint that it is very likely that
they are starburst galaxies at redshift higher than 0.5 (see La Franca
et al.  2004).

\subsection{The X-ray observations}

\begin{figure}
   \centering

   \includegraphics[angle=0,width=9cm]{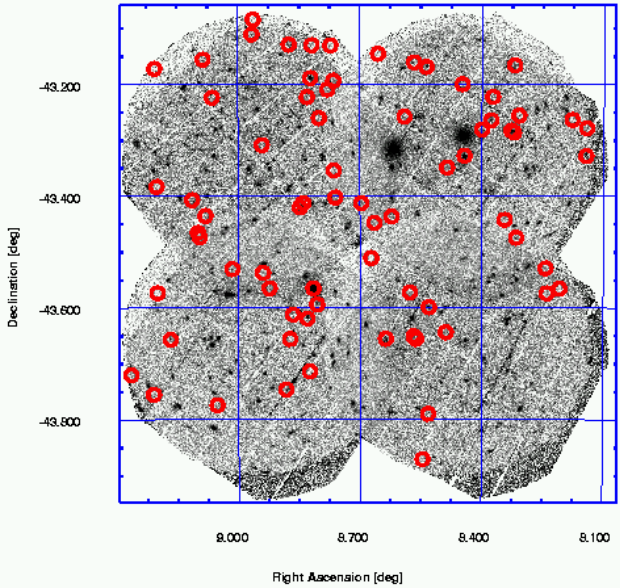}
      \caption{
                Mosaic of the four XMM-Newton pointings in the ELAIS-S1 field
        obtained by adding up the images of the PN and MOS cameras in
        the energy range 0.5-10 keV (Puccetti et al. 2006). The circles are the
position of the 72 15$\mu$m ELAIS extragalactic sources.
              }
         \label{FieldX}
   \end{figure}

The region S1-5 of the ELAIS was observed with the XMM-Newton
telescope. The observations and data reduction are described by
Puccetti et al. (2006). The region was observed by four partially
overlapping pointings, covering a total area of $\sim$0.6 deg$^2$ (see
Figure \ref{FieldX}). The observations were carried out from May to
July 2003, and each pointing lasted 84-100 Ksec. Source detection was
performed on Co-added PN+MOS1+MOS2 images accumulated in four energy
bands: 0.5--10 keV, 0.5--2 keV, 2--10 keV, and 5--10 keV. In total 478
sources were detected with a significance level\footnote{This
  corresponds to about one spurious source over each of the four
  XMM-Newton fields.} of 2$\times$10$^{-5}$. The flux limits are
5.5$\times$10$^{-16}$ erg cm$^{-2}$ s$^{-1}$ in the 0.5--2 keV band,
2$\times$10$^{-15}$ erg cm$^{-2}$ s$^{-1}$ in the 2--10 keV band, and
4$\times$10$^{-15}$ erg cm$^{-2}$ s$^{-1}$ in the 5--10 keV band.

%

The source counts in each camera were obtained using the events files
in the energy range 0.5-10 keV for the PN and 0.3-10 keV for the
MOS. The counts of the two MOS cameras were eventually combined.  The
counts of each source were extracted in a circular region with a
radius in the range $20\arcsec$--$30\arcsec$. In general, the radius
value was chosen so that the S/N ratio was roughly optimized, but this
choice was limited in a few cases by the presence of nearby sources
or by a peculiar position of the source on the detector, for example
close to a gap in the CCD array. In some cases the source was
detected, and the corresponding counts extracted, only in either the
PN or in one or both of the MOS cameras, because PN and MOS do not
cover exactly the same sky regions and the position of the gaps
differs in the PN, MOS1, and MOS2 CCD arrays.

The background counts for each source were extracted from the nearest
source-free region. In doing so, areas near gaps in the CCD array and
near the edge of the telescope field of view have been excluded, as
well as regions containing hot pixels and other CCD cosmetic defects.
 
The ancillary response files were generated for each source by means
of the tool {\sc arfgen}, in order to properly correct for
energy-dependent vignetting and point spread function. The response
matrix files were generated with the tool {\sc rmfgen} (SAS
6.1.0\footnote{
  http://xmm.vilspa.esa.es/external/xmm\_sw\_cal/sas\_frame.shtml}).

The spectral counts, when higher than about 120, were first
accumulated in energy bins with 20 counts each, from 0.3 keV to 10 keV
in the MOS and from 0.5 to 10 keV in the PN. They were then fitted,
using XSPEC (version 11.3.1) and the $\chi^2$ statistic with a model
comprising, in addition to the known galactic absorption
($\sim$2.76$\times$10$^{20}$ cm$^{-2}$): (1) a power law, with two
parameters, normalization and photon spectral index ${\Gamma}$; (2)
the absorption N$_H$ at the redshift of the source. Sometimes a thin
plasma model with the temperature as a parameter was added.  In the
case of degrees of freedom lower than 12, a fixed photon index
$\Gamma$=1.8 was assumed. When both PN and MOS data were available,
their relative normalization MOS/PN was left free to vary between 0.8
and 1.2. This interval was conservatively chosen wider than applicable
on-axis, because a fully reliable inter-calibration is still lacking
for sources off-axis . When the spectral counts were lower than about
120, the C statistic (Cash 1979) was used instead, as implemented in
XSPEC (Arnaud 2003 \footnote{K.A.  Arnaud, 2003, "XSPEC
  User Guide for version 11.3"\\
  http://heasarc.gsfc.nasa.gov/docs/software/lheasoft/xanadu/\\
  xspec/manual/manual.html}) after background subtraction (see
Alexander et al. 2003a for a similar procedure) and with 5 counts in
each energy bin (the latter choice was made only for convenience; it
does not impair the correct use of the embedded statistics when using
the above-mentioned XSPEC implementation).

\section{X-ray properties of the 15$\mu$m sources}

   We wish to use the X-ray observations to identify possible
   AGN activity among those 15$\mu$m galaxies whose optical spectroscopic
   identification did no show any AGN signatures. We have thus
   chosen a lower limit intrinsic un-absorbed X-ray luminosity of
   $\sim$10$^{42}$ erg s$^{-1}$ in order to classify a source as an AGN.
   Indeed higher luminosities are typical of AGN activity (see e.g.
   Ranalli, Comastri \& Setti 2003; Maiolino et al. 2003), while the
   most X-ray-luminous star-forming galaxy known, NGC 3256, has a
   total X-ray luminosity of L$_X$ $\simeq$ 8 $\times$ 10$^{41}$ erg
   s$^{-1}$ (Lira et al. 2002).
   
In total, 13 out of the 72 XMM observed ELAIS 15$\mu m$ sources were
detected in the X-ray band. Table 1 shows the X-ray and optical
spectral characteristics for each X-ray detected source. For the
remaining 59 X-ray undetected ISO sources, three sigma-confidence
level upper limits have been measured in the 0.5--10 keV and 2--10 keV
energy bands. Table 2 shows a summary for all 72 ELAIS sources of the
MIR, optical and X-ray fluxes and luminosities. The redshift and
optical spectroscopic classification is given when available. The
X-ray luminosities of the detected sources have been corrected for
absorption.

   \begin{figure}
   \centering
   \includegraphics[angle=-90,width=9cm]{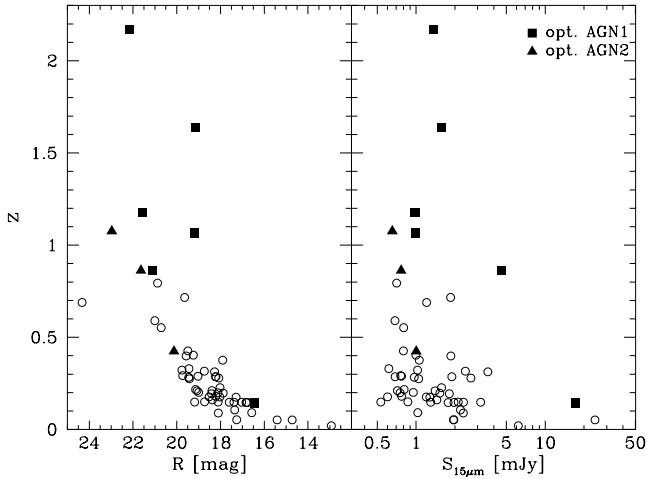}
      \caption{{\it Left}. Redshift versus R-band magnitude distribution of the 
15 $\mu$m sources.
Filled squares and triangles are optically classified AGN1 and AGN2, respectively. Open
circles are optically classified starburst galaxies. {\it Right}. Redshift versus
15 $\mu$m flux distribution of the  15 $\mu$m 
sources. Symbols are as in the previous panel.
              }
         \label{FluxZ}
   \end{figure}
%

   Twelve of the 13 X-ray detected sources have an optical and
   spectroscopic identification, while the X-ray detected source
   C15\_J003315-432829 has an optical counterpart with R=21.61 but no
   optical spectroscopic identification. Figure \ref{FluxZ} shows the
   redshift of the spectroscopically identified ELAIS sources as a
   function of the optical or MIR fluxes. As it is possible to see,
   all sources with R-band magnitude fainter than R=20.5 have redshift
   higher than 0.45. Then, as discussed by La Franca et al. (2004),
   this (not classified) source (C15\_J003315-432829) is likely to be
   a starburst galaxy or an AGN2 at redshift higher than 0.5.
   According to this guess, a lower limit of 10$^{43.4}$ erg s$^{-1}$
   on the X-ray luminosity is derived. This result allowed us to
   classify this source as an AGN.  Anyhow, for the sake of our
   analyses, even an assumed redshift as low as 0.2 would imply an
   X-ray luminosity higher than 10$^{42}$ erg s$^{-1}$, typical of
   AGN activity, and does not change our conclusions.
   
Four out of the six previously optically-classified AGN1 were detected
in the X-ray band and have 2--10 keV luminosities higher than
10$^{43.2}$ erg s$^{-1}$. Their X-ray spectral shape confirms this
classification, although two AGN1, C15\_J003330-431553 and
C15\_J003640-433925, show absorbed spectra with column densities of
logN$_H\sim$23 cm$^{-2}$.  This result is not so unusual for AGN1 (see
La Franca et al. 2005; and references therein). The remaining two
undetected AGN1 and the three (previously optically classified) AGN2
have 2--10 keV X-ray-luminosity upper limits of
$\sim$10$^{43.2}$-10$^{44.0}$ erg s$^{-1}$, which are still consistent
with the typical X-ray luminosity of un-absorbed AGNs
(10$^{42}$-10$^{46}$ erg s$^{-1}$).

   \begin{figure}
   \centering
   \includegraphics[angle=-90,width=9cm]{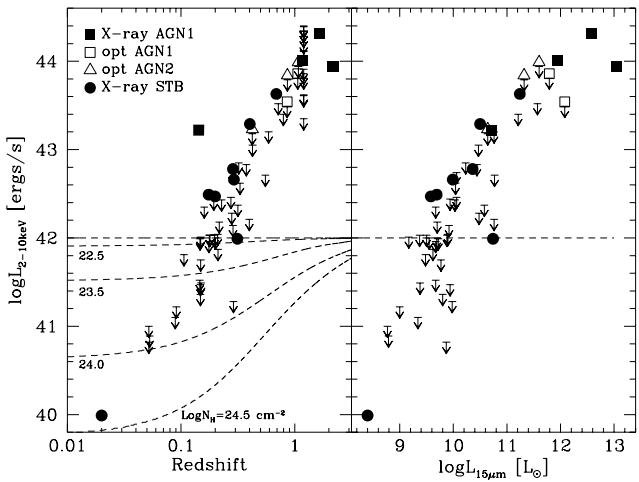}
  \caption{{\it Left}. 2-10 keV X-ray
    luminosity versus redshift. Squares are AGN1, triangles are
    optically classified AGN2, and circles are galaxies. Filled
    symbols are X-ray detected sources. The dashed lines show our AGN
    luminosity threshold, and the expected observed X-ray luminosity as a
    function of absorption and redshift of an absorbed source having
    an intrinsic X-ray luminosity of 10$^{42}$ erg s$^{-1}$.  {\it
      Right}. 2-10 keV X-ray luminosity versus 15 $\mu$m luminosity
    of the ELAIS-S1-5 sources. Symbols are as in the previous panel.}
         \label{LLS}
   \end{figure}
%

   \begin{figure*}
   \centering
   \includegraphics[width=12cm,angle=-90]{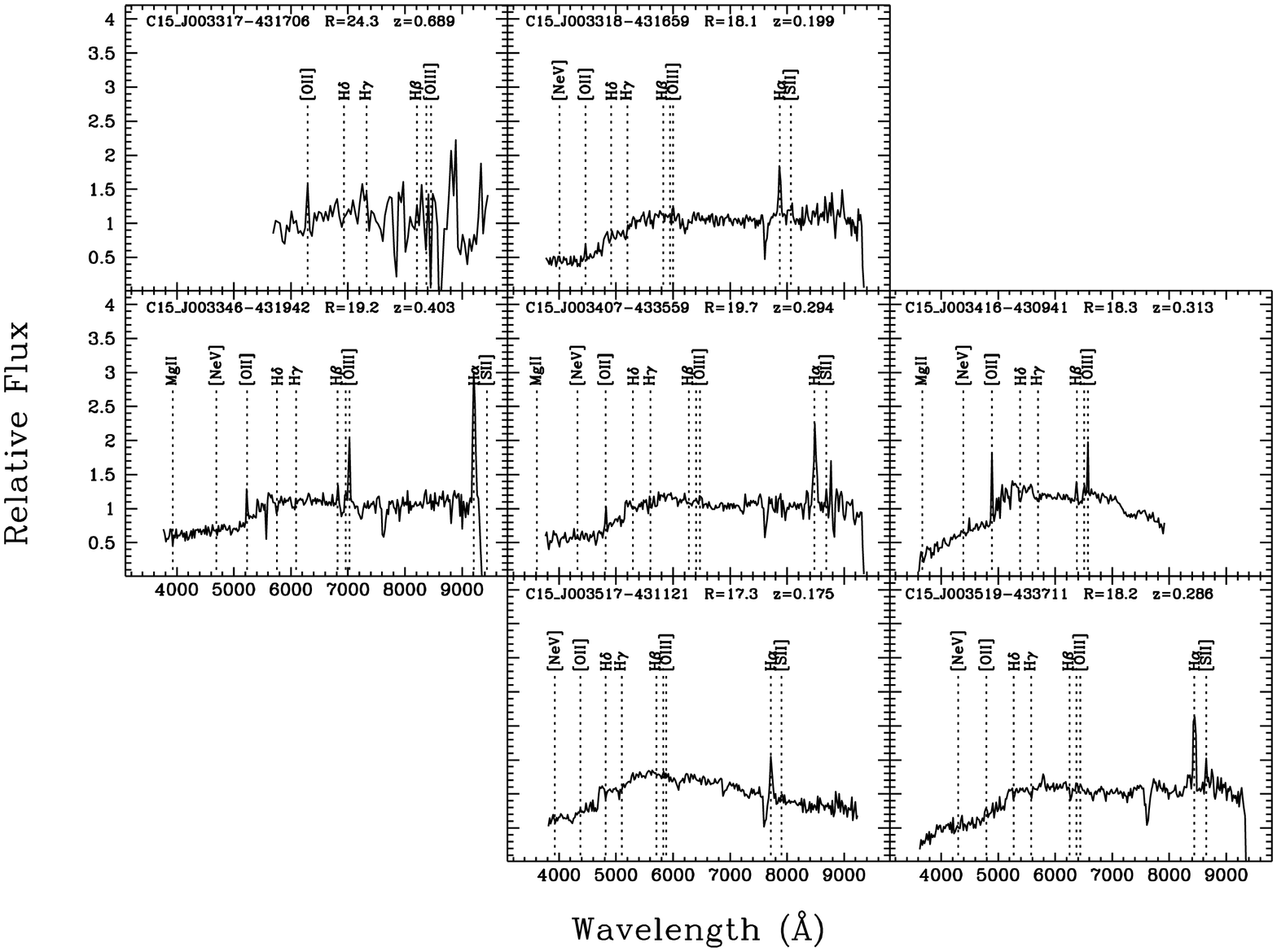}
\caption{Optical spectra of the seven objects showing AGN X-ray luminosity
and previously classified as no-AGN galaxies. The position of the typical
emission lines of AGN (even if not observed) are shown.}

         \label{Spettri}
   \end{figure*}
%

   All the 9 remaining X-ray detected sources have the optical spectra
   typical of starburst activity. However, eight of them have X-ray
   luminosities typical of AGN activity: five showed absorbed
   (logN$_H\geq $22 cm$^{-2}$) X-ray spectra, thus reinforcing the
   classification as AGN2, while the other three sources have X-ray
   spectra consistent with an unabsorbed power law. Finally, the last
   source (C15\_J003432-433922) has the X-ray luminosity typical of
   starburst activity ($\sim$10$^{40}$ erg s$^{-1}$).  This is also
   confirmed by its X-ray spectrum, where a thin plasma model (MEKAL
   with KT=0.60$^{+1.11}_{-0.35}$) has necessarily been added to the
   power-law model.

   There are 14 sources without X-ray detections and without measured
   redshift. According to either their R-band optical magnitudes or
   their R-band magnitude limits (R$>$24.5), most of these sources
   should lie at redshifts higher than 0.5 (see discussion in La Franca
   et al. 2004).  In order to derive an upper limit on their X-ray
   luminosities, we made the conservative assumption that these
   sources lie at redshift 1.2. According to this assumption, the
   corresponding 2--10 keV unabsorbed X-ray luminosity upper limits
   are always greater than 10$^{43}$ erg s$^{-1}$ (see Fig.
   \ref{LLS}), and then no conclusion on their nature can be derived
   from the X-ray observations.

Figure \ref{LLS} shows the 2--10 keV unabsorbed luminosity as a
function of the redshift and MIR luminosities.  For 59 sources, we
have only upper limits on the X-ray fluxes and, then, on the {\it
un-absorbed} X-ray luminosities (using the measured redshifts or, when
not available, a conservative estimate of $z$=1.2). However these
estimates do not take the effects of possible X-ray absorption into
account . This effect is more relevant at low redshifts (z$<$0.5-1)
because the absorption mainly affects the rest-frame, lower-energy
photons that are instead no longer observable in the 2-10 keV band at
higher redshift. In Fig. \ref{LLS} (left) the expected observed X-ray
luminosity is shown as a function of absorption and the redshift of an
absorbed source having an intrinsic X-ray luminosity of 10$^{42}$ erg
s$^{-1}$. We can expect 50\% less emission at redshift $z\sim0.5$ for
a Compton-thin source with a column density of LogN$_H$=23.5
cm$^{-2}$. In this case a $\sim$10$^{41.7}$ erg s$^{-1}$ absorbed
luminosity would be observed. The only solution for detecting such low
redshift-absorbed AGN would be to obtain deeper X-ray observations in
order to analyze their X-ray spectra and separate X-ray luminous
star-forming galaxies from absorbed AGN. As expected, the situation is
even worst for Compton-thick (LogN$_H$$>$24 cm$^{-2}$), absorbed AGN:
in this case the fluxes would be reduced by more than a factor of
ten. We thus conclude that our X-ray observations are only able to
detect a relevant, but nevertheless incomplete, fraction of the
Compton-thin AGN. Then our attempt to identify the population of AGN
missed by the optical spectroscopic identifications is partly
inefficient, limited to un-absorbed, or Compton-thin, absorbed AGN.
Therefore, the densities of AGN derived in the next sections are only
to be considered lower limits.

   \begin{figure}[t]
   \centering
   \includegraphics[width=9cm]{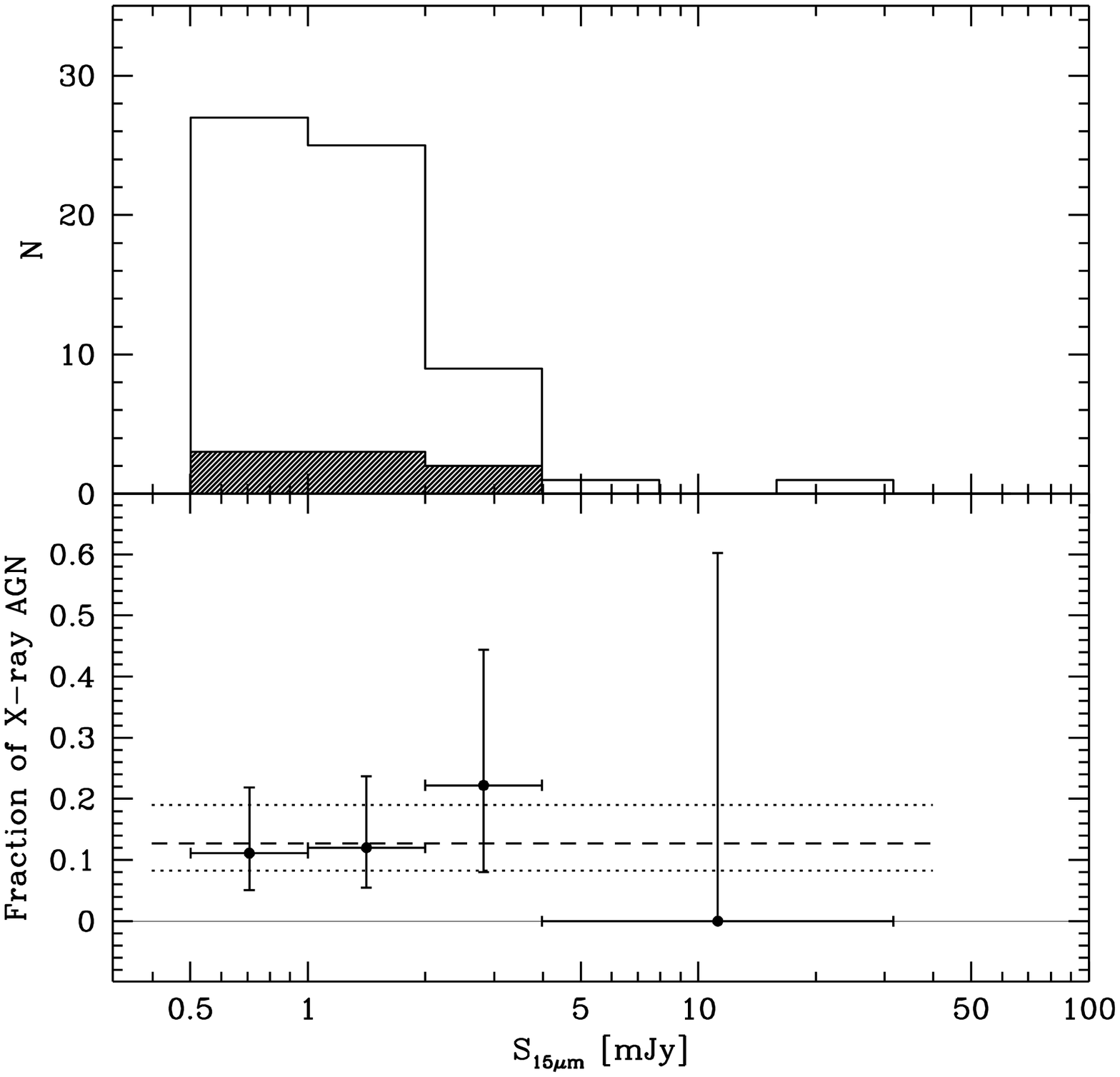}
\caption{{\it Up}. Flux distribution of the 63 ELAIS-S1-5 sources that were
  not optically classified as AGN. The shaded histogram shows the flux
  distribution of the eight sources that showed AGN activity in the
  X-ray band. {\it Down}. Fraction of the X-ray active AGN among the
  galaxies as a function of the 15$\mu m$ flux. The long dashed line
  shows the adopted average constant value of 12.7\%, while the dotted
  lines are 1$\sigma$ confidence-level uncertainties.}
         \label{Fract}
   \end{figure}
%

\section{Discussion}

\subsection{The newly classified AGN}

We observed 72 ELAIS sources with XMM-Newton. Nine sources, previously
classified in the optical as AGN (6 AGN1 and 3 AGN2) showed X-ray
properties (either detections or upper limits) in agreement with their
optical classification.  Eight out of the 63 previously classified (or
bona fide) starburst galaxies instead showed X-ray luminosities
typical of AGN activity.  As our optical spectroscopic studies were
sensitive enough to detect broad emission lines (see La Franca et
al. 2004), we classified them as new AGN2. As discussed in the
previous section, the object C15\_J003315-432829 has no spectroscopic
identification, but its X-ray flux is so bright that will have an X-ray
luminosity higher than 10$^{42}$ erg s$^{-1}$ at redshifts higher than
z=0.2. We inspected the optical spectra of the remaining seven new
AGN2 (see Fig. \ref{Spettri}).  C15\_J003317-431706 was observed
after the publication of the spectroscopic catalogue by La Franca et
al. (2004) and is an R=24.3 galaxy at z=0.689 observed by VIMOS at VLT
(see \S\ 2.1). It has a noisy spectrum, just showing a clear
[OII]$\lambda$3727 emission line, which hampers a more detailed
classification.  According to the classification scheme used by La
Franca et al. (2004; see also Dressler et al.  1999; Poggianti et
al. 1999), the remaining six galaxies are: two e(a) dust-enshrouded
starburst galaxies, three e(c) star-forming spirals, and one k(e)
post-starburst galaxy with emission lines.  C15\_J003346-431942, which
was classified as an e(a) galaxy, has the line ratios typical of a
LINER, but could be compatible with an AGN2 classification if 10\%
errors are taken into account. C15\_J003318-431659, which was
classified as an e(c) galaxy (showing no detectable H$\beta$ emission
or absorption) has a limit on the [OIII]5007/H$\beta$ ratio which is
compatible with an AGN2. On the contrary, even taking the line ratio
uncertainties into account, the remaining five objects do not show any
signature in the optical spectrum compatible with an AGN activity.

As shown in Fig. \ref{Fract}, according to the large statistical
uncertainties, there is no significant dependence on the 15$\mu m$
flux by the fraction of the newly X-ray classified AGN2. We can then
conclude that at least a constant fraction of 12.7$^{+6.3}_{-4.4}$\%
of the previously optically-classified starburst galaxies are instead
AGN2.

According to our X-ray luminosities' upper limits or measurements, all
galaxies with L$_{15}$$<$ 10$^{9.5}$ L$_\odot$ are unlikely to harbor
an AGN (see Fig. \ref{LLS}). In contrast, all the newly classified
AGN2 have IR luminosities in the range 10$^{9.5}$ L$_\odot$ $<$
L$_{15}$$<$ 10$^{11}$ L$_\odot$, exactly in the region of the AGN2 LF
where a flattening has been observed (Matute at al. 2006). This result
confirms the hypothesis of Matute at al. (2006) that the observed
flattening of the AGN2 LF at redshift $\sim$0.35 was due to an
incompleteness in the AGN identification. As previously pointed out by
La Franca et al. (2004), the origin of this incompleteness has to be
ascribed to the fact that a relevant fraction of AGN2 do not show
clear AGN signatures in their optical spectra.
 
The more IR luminous objects in our sample are AGN1, which show
15$\mu$m luminosities higher than 10$^{11.5}$ L$_\odot$. As discussed
by Matute et al. (2006), the absence of AGN2 in this luminosity range
is mainly due to selection effects.  The AGN1 SEDs have a fairly
constant infrared to optical ratio, while for AGN2 the
infrared-to-optical ratio increases with the increase in the
luminosity (Log(L$_{15}$/L$_R$) $\propto$ 0.5 LogL$_{15}$). As a
consequence, AGN2 have on average fainter optical counterparts than
AGN1 at high luminosities and then miss the spectroscopic
identification (see Fig. 1 in Matute et al. 2006).

\subsection{Revised AGN 15$\mu m$ counts}

\begin{figure*}
   \centering
   \includegraphics[width=12cm,angle=-90]{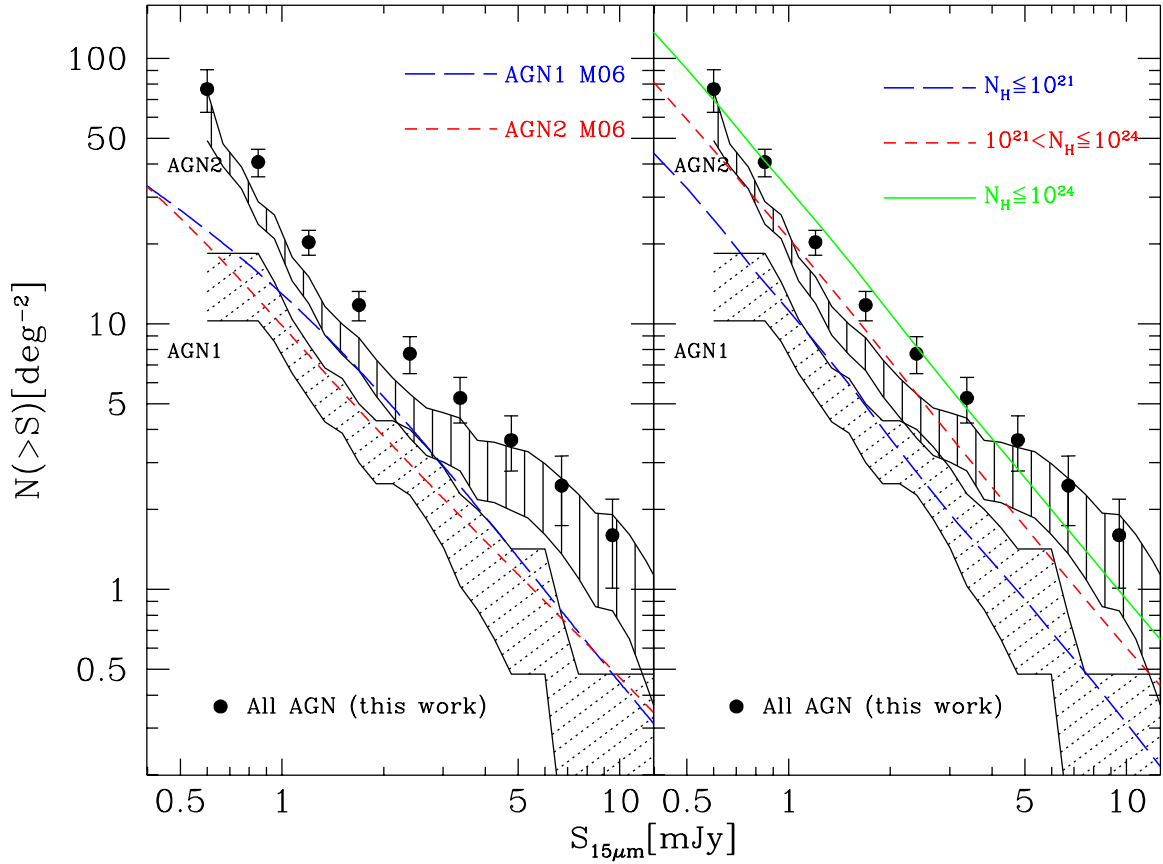}
   \caption{ Revised AGN1 and (Compton thin) AGN2 integral counts at
     15$\mu m$.  Filled circles are the total (AGN1+AGN2) counts. {\it
       Left.} The short and long dashed lines are the predicted AGN1
     and AGN2 counts, respectively, by Matute et al. (2006). {\it
       Right.} The short and long dashed lines are the predicted
     un-absorbed and Compton thin counts, respectively, by Silva et
     al. (2004) on the basis of the X-ray AGN luminosity function and
     semi-empirical SEDs (linking the X--ray to the infrared). The
     continuous line is the prediction of the total AGN counts
     (including Compton-thin absorbed sources) by Silva et
     al. (2004).}

         \label{Counts15}
   \end{figure*}

According to this above new classification, we have then corrected the
integral 15$\mu m$ counts, previously derived from the optical
spectroscopic classification of the ELAIS-S1 sources by La Franca et
al. (2004).  According to the estimates described in the previous
section, we have assumed that 12.7\% of all ELAIS-S1 optically
classified or bona fide starburst galaxies harbor an AGN2. The results
are presented in Table 3 and shown in Fig. \ref{Counts15}. The newly
derived counts are compared in Fig. \ref{Counts15} (left) with the
predictions from the AGN1 and AGN2 luminosity functions computed by
Matute et al. (2006), and in Fig. \ref{Counts15} (right) with the AGN
counts predicted by Silva et al. (2004) on the basis of the X-ray AGN
luminosity function (LF) and semi-empirical SEDs (linking the X--ray
to the infrared). It turns out that the AGN1 counts agree, within the
uncertainties, with the counts predicted by both the AGN1 LF of Matute
et al. (2006) and the un-absorbed (N$_H$$\leq$ 10 $^{21}$ cm$^{-2}$)
AGN of Silva et al. (2004). Conversely, the new AGN2 counts are, as
expected, about a factor of 2--3 higher than predicted by the AGN2 LF
of Matute et al. (2006). At fluxes larger than 0.6 mJy, the density of
AGN2 is about 60 deg$^{-2}$, while Matute et al. (2006) predict a
density of about 20 deg$^{-2}$.  On the other hand, the predicted
count of absorbed, Compton-thin AGN (10$^{21}$$<$N$_H$$<$10$^{24}$
cm$^{-2}$) by Silva et al. (2004) are consistent with our estimates of
the AGN2 counts.

All the extragalactic sources, with 15 $\mu m$ fluxes brighter than
0.6 mJy, produce a 15 $\mu m$ background of $\nu$I($\nu$)= 0.343 nW
m$^{-2}$ sm$^{-1}$, while the AGNs produce $\nu$I($\nu$)= 0.046 nW
m$^{-2}$ sm$^{-1}$ ($\sim$13\% contribution to the extragalactic
background produced at fluxes brighter than 0.6 mJy).

\begin{table}
\setcounter{table}{2}
\caption{ AGN1 and AGN2  integral counts at 15 $\mu$m}
\label{table3}
\begin{tabular}{l c c}
\hline \hline 
S$_{15}$ & AGN1 & AGN2 \\
 mJy & N($>$S) deg$^{-2}$ &  N($>$S) deg$^{-2}$ \\
\hline

  0.60 &   14.4$\pm$   4.1 &   62 $\pm$  13 \\ 
  0.76 &   14.4$\pm$   4.1 &   35.6 $\pm$   3.4  \\
  0.95 &   11.5$\pm$   2.9 &   23.3 $\pm$   2.4  \\
  1.20 &    6.8$\pm$   1.6 &   13.5 $\pm$   1.6 \\ 
  1.51 &    5.1$\pm$   1.3 &    8.8 $\pm$   1.2 \\ 
  1.90 &   3.41$\pm$   0.91 &   6.3 $\pm$   1.0  \\
  2.39 &   3.14$\pm$   0.87 &   4.58$\pm$   0.86  \\
  3.01 &   2.15$\pm$   0.72 &   3.82$\pm$   0.81  \\
  3.79 &   1.42$\pm$   0.58 &   2.91$\pm$   0.73  \\
  4.77 &   0.95$\pm$   0.47 &   2.70$\pm$   0.72  \\
  6.00 &   0.95$\pm$   0.47 &   2.29$\pm$   0.68  \\
  7.55 &   0.24$\pm$   0.24 &   1.69$\pm$   0.59  \\
  9.51 &   0.24$\pm$   0.24 &   1.37$\pm$   0.54  \\

\hline
\end{tabular}
\end{table}

%

\section{Conclusions}

We have used XMM-Newton observations of the 15 $\mu$m sample of the
ELAIS-S1-5 area to disentangle AGN activity among previously
spectroscopically classified starburst galaxies. We find that at least
a fraction of 13$\pm$6\% (not dependent on the flux) of the previously
spectroscopically classified starburst galaxies are instead bona fide
(Compton thin) AGN2. The fraction of newly identified AGN2 is not
large enough to significantly change the results previously found by
Pozzi et al. (2004) for the evolution of the LF of the starburst
galaxies obtained by using the ELAIS sample.  On the other hand, as
the density of galaxies is about an order of magnitude higher than the
density of AGN at 15 $\mu$m fluxes of $\sim$1 mJy, the new density of
AGN2 is significantly increased by a factor of $\sim$2-3. This result
agrees with the analysis of Matute et al. (2006), who ascribe the
observed flattening of the AGN2 LF at redshifts $\sim$0.5 to an
incompleteness of the optical spectroscopic identification techniques.

Here we wish to focus on the two results: a) that more than half of
the AGN2 could miss their classification if only the optical spectra
were used and b) that, according to our analysis and predictions
derived from the hard X-ray luminosity function, the surface density
of (Compton thin) AGN2 at fluxes larger than 0.6 mJy is in the range
50-70 deg $^{-2}$.  These are the first estimates of separated AGN1
and AGN2 counts in the MIR. This result corresponds to a fraction of
about 24\% (17/72) of AGN (AGN1 plus AGN2) among all the extragalactic
sources at 15$\mu$m fluxes larger than 0.6 mJy ($\sim$13\%
contribution to the extragalactic background produced at fluxes
brighter than 0.6 mJy).

According to our analysis, the ongoing studies of the density and
evolution of absorbed/reddened AGN (AGN2) among the {\it Spitzer}
sources, if only based on optical spectroscopic identifications, would
miss about half of the true AGN2.  A more efficient approach would
require MIR color-selection criteria, or better if complemented with
deep hard X-ray observations (see e.g. Barmby et al.  2006; Donley et
al.  2007; Fiore et al. 2007; Daddi et al. 2007).

\begin{acknowledgements}
  
  We thank Laura Silva for discussions and for having provided the
  predictions of the 15$\mu$m AGN counts in machine-readable format.
  We thank the anonymous referee for the useful comments.  This
  research has been partially supported by ASI, INAF, and MIUR grants.

\end{acknowledgements}


\setcounter{table}{0}
\begin{landscape}
\begin{table}
\caption{X-ray detected sources}
\label{table2}
\begin{tabular}{c c c c c c l l l c r r c r r c l }
\hline \hline 
Name & $\alpha_{MIR}$ & $\delta_{MIR}$ &$\alpha_{X}$ & $\delta_{X}$ &$\Delta$ & $\Gamma$ & KT & N$_H$ & $\chi^2/dof$ & F$_{0.5-10}$ & F$_{2-10}$ & z &L$_{0.5-10}$ & L$_{2-10}$ & Cl$_O$ & Cl$_X$ \\
 (1) & (2) & (3) & (4) & (5) & (6) & (7) & (8) & (9) & (10) & (11) & (12)  & (13) & (14) & (15) & (16) & (17)\\
\hline

C15\_J003234-431940&  8.143084&-43.327194& 8.14298&-43.32731&0.5&$1.80$              & ...               &$0.00_{0.00}^{0.43} $& 6.8/10  & 2.28& 1.44&1.637&44.53&44.32& 1 & AGN1\\ 
C15\_J003315-432829&  8.315042&-43.475304& 8.31612&-43.47545&2.9&$2.37_{2.00}^{4.22}$& ...               &$8.0_{7.5}^{20.2}   $& 10.6/14& 1.81& 1.69&$>$0.5&$>$43.78&$>$43.40&...& AGN \\
C15\_J003317-431706&  8.321417&-43.285069& 8.31999&-43.28590&4.8&$1.80$              & ...               &$4.32_{1.21}^{1.66} $& 2.0/4  & 2.49& 2.28&0.689&43.68&43.63& 5 & AGN2 \\
C15\_J003318-431659&  8.326542&-43.284195& 8.32648&-43.28473&1.9&$1.80$              & ...               &$3.39_{0.98}^{1.25} $&27.3/27 & 2.38& 2.24&0.199&42.68&42.47& 5 & AGN2 \\
C15\_J003330-431553&  8.376667&-43.265110& 8.37693&-43.26542&1.3&$1.80$              & ...               &$~7.2_{~5.1}^{12.8} $&12.3/22 &0.43&0.34&2.170&44.21&43.94& 1 &  AGN1 \\
C15\_J003346-431942&  8.442875&-43.328835& 8.44301&-43.32917&1.3&$1.49_{0.10}^{0.16}$& ...               &$0.00_{0.00}^{0.10} $&60.9/61 & 5.44& 3.97&0.403&43.43&43.29& 3 & AGN2 \\
C15\_J003407-433559&  8.530666&-43.599583& 8.53035&-43.59983&1.2&$2.02_{0.60}^{1.07}$& ...               &$1.45_{0.97}^{1.68} $&26.5/31 & 1.89& 1.54&0.294&42.94&42.66& 5 & AGN2 \\
C15\_J003416-430941&  8.566916&-43.161446& 8.56900&-43.16218&6.1&$1.80$              &...                &$0.02_{0.02}^{1.03} $& 3.0/4  & 0.32& 0.05&0.313&42.20&42.00& 3 & AGN2 \\
C15\_J003432-433922&  8.635791&-43.656029& 8.63476&-43.65776&6.8&$1.80$             &$0.60_{0.35}^{1.11}$&$0.00_{0.00}^{0.12} $& 8.1/8  & 2.01& 1.07&0.020&40.28&39.99& 6 & STB \\
C15\_J003515-433356&  8.815042&-43.566029& 8.81537&-43.56558&1.8&$2.68_{0.00}^{0.04}$& ...               &$0.00_{0.00}^{0.00} $&285/160 &86.24&27.62&0.143&43.75&43.22& 1 & AGN1 \\
C15\_J003517-431121&  8.821792&-43.189335& 8.82196&-43.18954&0.9&$1.90_{0.10}^{0.22}$& ...               &$0.00_{0.00}^{0.04} $&52.7/56 & 6.15& 3.66&0.175&42.73&42.49& 6 & AGN2 \\
C15\_J003519-433711&  8.830126&-43.619667& 8.82976&-43.61887&3.0&$1.80$              & ...               &$2.04_{0.79}^{1.11} $&13.7/11 & 2.52& 2.22&0.286&42.98&42.78& 5 & AGN2 \\
C15\_J003640-433925&  9.168375&-43.657333& 9.16840&-43.65727&0.2&$1.80$              & ...               &$10.8_{~4.7}^{13.5} $&28.1/26 & 1.47& 1.34&1.181&44.20&44.01& 1 & AGN1 \\

\hline
\multicolumn{16}{l}{Notes.-- The columns are as follows: (1) ELAIS ID
  name; (2, 3) Right Ascension and Declination at 15 $\mu$m (deg); (4, 5) Right Ascension and Declination in the 0.5-10 keV band;}\\

\multicolumn{16}{l}{ (6) distance of the MIR and X-ray sources (arcsec); (7) photon spectral index; (8) Temperature of the MEKAL model (keV); (9) N$_H$ column densities
in 10$^{22}$ cm$^{-2}$ units; }\\

\multicolumn{16}{l}{ (10) $\chi^2$ and degrees of freedom of the XSPEC fit; (11, 12 ) 0.5-10 keV and 2-10 keV band fluxes in 10$^{-14}$ erg/s/cm$^2$ units;
(13) redshift; }\\

\multicolumn{16}{l}{ (14, 15) log of the 0.5-10 keV band and 2-10 keV band luminosity in erg/s units; (16)
Optical spectroscopic class as in La Franca
et al. (2004) (1=AGN1, 2=AGN2, 3=e(a) gal,}\\

\multicolumn{16}{l} {  4=e(b) gal, 5=e(c) gal, 6=k(e) gal, 7=k gal); (17) Optical plus X-ray final classification.}\\

\end{tabular}
\end{table}
\end{landscape}


\onecolumn
\normalsize

\setcounter{table}{1}
\begin{center}
\setlength\LTleft{0pt}
\setlength\LTright{0pt}
\begin{longtable}{ c c c c c c c r r r r }
\caption{X ray properties of the 15$\mu m$ sources}
\label{table1}
\\
\hline \hline 
Name & F$_{15}$ & R & $z$ & Cl & L$_{15}$ & L$_R$ & F$_{0.5-10}$ & F$_{2-10}$ & L$_{0.5-10}$ & L$_{2-10}$\\
 (1) & (2) & (3) & (4) & (5) & (6) & (7) & (8) & (9) & (10) & (11)\\
\hline
\endfirsthead

\caption{Continued from previous page}\\
\hline\hline
Name & F$_{15}$ & R & $z$ & Cl & L$_{15}$ & L$_R$ & F$_{0.5-10}$ & F$_{2-10}$ & L$_{0.5-10}$ & L$_{2-10}$\\
 (1) & (2) & (3) & (4) & (5) & (6) & (7) & (8) & (9) & (10) & (11)\\
\hline
\endhead

\hline
\multicolumn{11}{l}{Continued in next page}\\
\endfoot

\multicolumn{11}{l}{Notes.-- The columns are as follows: (1) ELAIS ID
  name; (2) 15$\mu$m flux in mJy units; (3) R-band magnitude;}\\

\multicolumn{11}{l} {(4) redshift; (5) Optical spectroscopic class as in La Franca
et al. (2004) (1=AGN1, 2=AGN2, 3=e(a) gal, 4=e(b) gal,}\\

\multicolumn{11}{l} { 5=e(c) gal, 6=k(e) gal, 7=k gal); (6) 15$\mu$m luminosity in solar units;
(7) R-band luminosity in solar units;}\\

\multicolumn{11}{l} { (8) 0.5-10 keV band flux in 10$^{-14}$ erg/s/cm$^2$ units; 
(9) 2-10 keV band flux in 10$^{-14}$ erg/s/cm$^2$ units;}\\

\multicolumn{11}{l} { (10) log of the 0.5-10 keV band luminosity in erg/s units;
(11) log of the 2-10 keV band luminosity in erg/s units.}\\

\endlastfoot
ELAISC15\_J003234-431642&  1.18&19.12&   ...&0&  ...&  ...&$<$ 0.61&$<$ 1.62&$<$43.86&$<$44.28\\ 
ELAISC15\_J003234-431940&  1.58&19.15& 1.637&1&12.57&12.29&  2.28&  1.44& 44.53& 44.32\\ 
ELAISC15\_J003242-431548&  0.71&20.88& 0.794&4&11.21&11.17&$<$ 0.43&$<$ 0.96&$<$43.05&$<$43.40\\ 
ELAISC15\_J003249-433201&  1.65&23.70&   ...&0&  ...&  ...&$<$ 0.49&$<$ 1.36&$<$43.76&$<$44.21\\ 
ELAISC15\_J003257-433426&  0.77&18.21& 0.180&5& 9.50&10.22&$<$ 0.72&$<$ 1.23&$<$41.80&$<$42.03\\ 
ELAISC15\_J003258-433145&  0.77&19.03& 0.289&5& 9.98&10.41&$<$ 0.12&$<$ 0.08&$<$41.49&$<$41.28 \\ 
ELAISC15\_J003314-431522&  1.40&19.08& 0.210&6& 9.90&10.03&$<$ 0.65&$<$ 0.90&$<$41.90&$<$42.04 \\ 
ELAISC15\_J003315-432829&  2.50&21.61&$>$0.5&0&  ...&  ...&  1.81&  1.69&$>$43.78&$>$43.40\\ 
ELAISC15\_J003316-430959&  1.53&17.88& 0.197&6& 9.88&10.44&$<$ 1.03&$<$ 1.01&$<$42.04&$<$42.03\\ 
ELAISC15\_J003317-431706&  1.21&24.33& 0.689&5&11.24& 9.51&  2.49&  2.28& 43.68&43.63\\ 
ELAISC15\_J003318-431659&  0.75&18.12& 0.199&5& 9.58&10.36&  2.38&  2.24& 42.68&42.47\\ 
ELAISC15\_J003322-432633&  2.41&18.74& 0.316&3&10.58&10.63&$<$ 0.39&$<$ 0.77&$<$42.08&$<$42.37\\ 
ELAISC15\_J003329-431322&  0.72&18.39& 0.211&6& 9.62&10.31&$<$ 0.11&$<$ 0.60&$<$41.12&$<$41.87\\ 
ELAISC15\_J003330-431553&  1.36&22.16& 2.170&1&13.05&12.76&  0.43&  0.34& 44.21& 43.94\\ 
ELAISC15\_J003335-431653&  0.87&18.73& 0.150&5& 9.38& 9.82&$<$ 0.50&$<$ 0.52&$<$41.47&$<$41.49\\ 
ELAISC15\_J003346-431942&  1.00&19.24& 0.403&3&10.50&10.74&  5.44&  3.97& 43.43& 43.29\\ 
ELAISC15\_J003347-431201&  0.81&19.14& 0.217&5& 9.70&10.04&$<$ 0.58&$<$ 1.13&$<$41.89&$<$42.18\\ 
ELAISC15\_J003356-432058&  2.33&17.03& 0.148&5& 9.80&10.49&$<$ 0.33&$<$ 0.40&$<$41.28&$<$41.36\\ 
ELAISC15\_J003357-433839&  0.81&21.87&   ...&0&  ...&  ...&$<$ 0.30&$<$ 0.35&$<$43.55&$<$43.61\\ 
ELAISC15\_J003401-430846&  1.95&14.72& 0.052&7& 8.77&10.41&$<$ 0.41&$<$ 1.58&$<$40.42&$<$41.00\\ 
ELAISC15\_J003407-433559&  0.76&19.73& 0.294&5& 9.99&10.15&  1.89&  1.54& 42.94& 42.66         \\ 
ELAISC15\_J003407-434725&  0.80&20.71& 0.552&3&10.77&10.61&$<$ 0.24&$<$ 0.46&$<$42.43&$<$42.71 \\ 
ELAISC15\_J003408-431011&  0.99&19.20& 1.065&1&11.79&11.75&$<$ 0.44&$<$ 1.39&$<$43.36&$<$43.86 \\ 
ELAISC15\_J003410-435217&  1.08&  ...&   ...&0&  ...&  ...&$<$ 0.13&$<$ 0.35&$<$43.19&$<$43.62 \\ 
ELAISC15\_J003414-433012&  0.65&22.80&   ...&0&  ...&  ...&$<$ 0.35&$<$ 1.08&$<$43.61&$<$44.11 \\ 
ELAISC15\_J003416-430941&  3.60&18.27& 0.313&3&10.74&10.80&  0.32&  0.05& 42.20& 42.00         \\ 
ELAISC15\_J003416-433905&  1.03&16.57& 0.091&5& 9.00&10.19&$<$ 0.80&$<$ 0.82&$<$41.21&$<$41.22 \\ 
ELAISC15\_J003417-433422&  0.53&19.18& 0.149&3& 9.17& 9.63&$<$ 0.40&$<$ 1.77&$<$41.37&$<$42.01 \\ 
ELAISC15\_J003421-431531&  3.16&17.38& 0.148&6& 9.94&10.35&$<$ 0.48&$<$ 0.51&$<$41.44&$<$41.47 \\ 
ELAISC15\_J003429-432614& 24.32&15.41& 0.052&5& 9.87&10.13&$<$ 0.75&$<$ 1.04&$<$40.68&$<$40.82 \\ 
ELAISC15\_J003432-433922&  6.19&12.93& 0.020&6& 8.40&10.25&  2.01&  1.07& 40.28& 39.99         \\ 
ELAISC15\_J003439-432654&  1.96&17.26& 0.053&6& 8.79& 9.41&$<$ 0.56&$<$ 1.17&$<$40.57&$<$40.89 \\ 
ELAISC15\_J003441-433041&  1.45&18.37& 0.160&6& 9.67&10.03&$<$ 4.57&$<$ 3.26&$<$42.49&$<$42.35 \\ 
ELAISC15\_J003447-432447&  0.65&22.97& 1.076&2&11.60&10.86&$<$ 0.98&$<$ 1.96&$<$43.71&$<$43.99 \\ 
ELAISC15\_J003502-432411&  1.58&18.02& 0.227&5&10.03&10.54&$<$ 0.90&$<$ 1.84&$<$42.12&$<$42.43 \\ 
ELAISC15\_J003503-432117&  1.77&16.83& 0.146&5& 9.67&10.55&$<$ 0.14&$<$ 0.60&$<$40.90&$<$41.52 \\ 
ELAISC15\_J003503-431138&  1.20&18.03& 0.176&5& 9.67&10.27&$<$ 0.61&$<$ 1.09&$<$41.71&$<$41.96 \\ 
ELAISC15\_J003505-430752&  1.03&19.76& 0.322&3&10.23&10.24&$<$ 1.67&$<$ 2.23&$<$42.73&$<$42.85 \\ 
ELAISC15\_J003507-431236&  0.60&18.52& 0.177&3& 9.37&10.08&$<$ 0.73&$<$ 1.26&$<$41.79&$<$42.03 \\ 
ELAISC15\_J003512-431540&  1.05&19.42& 0.275&6&10.06&10.19&$<$ 0.61&$<$ 1.27&$<$42.13&$<$42.46 \\ 
ELAISC15\_J003513-433540&  1.97&16.78& 0.147&6& 9.72&10.58&$<$ 1.70&$<$ 1.72&$<$41.98&$<$41.99 \\ 
ELAISC15\_J003515-433356& 17.32&16.43& 0.143&1&10.71&10.66& 86.24& 27.62& 43.75& 43.22         \\ 
ELAISC15\_J003516-430750&  0.96&24.88&   ...&0&  ...&  ...&$<$ 0.35&$<$ 0.55&$<$43.61&$<$43.81 \\ 
ELAISC15\_J003517-431121&  1.28&17.29& 0.175&6& 9.69&10.56&  6.15&  3.66& 42.73& 42.49         \\ 
ELAISC15\_J003517-434252&  1.30&17.60& 0.148&6& 9.55&10.26&$<$ 1.26&$<$ 1.73&$<$41.86&$<$42.00 \\ 
ELAISC15\_J003519-433711&  1.89&18.21& 0.286&5&10.36&10.72&  2.52&  2.22& 42.98& 42.78         \\ 
ELAISC15\_J003519-431325&  2.66&18.08& 0.279&6&10.48&10.74&$<$ 0.86&$<$ 0.83&$<$42.30&$<$42.28 \\ 
ELAISC15\_J003520-433645&  2.12&18.11& 0.149&3& 9.77&10.06&$<$ 0.59&$<$ 0.96&$<$41.54&$<$41.75 \\ 
ELAISC15\_J003521-432447&  2.32&18.10& 0.089&4& 9.34& 9.56&$<$ 0.40&$<$ 0.65&$<$40.89&$<$41.10 \\ 
ELAISC15\_J003523-432514&  0.97&19.45& 0.283&3&10.06&10.21&$<$ 1.34&$<$ 2.32&$<$42.51&$<$42.74 \\ 
ELAISC15\_J003529-433923&  0.58&  ...&   ...&0&  ...&  ...&$<$ 0.26&$<$ 0.62&$<$43.49&$<$43.86 \\ 
ELAISC15\_J003529-430746&  0.95&19.00& 0.201&3& 9.69&10.02&$<$ 0.71&$<$ 0.90&$<$41.90&$<$42.00 \\ 
ELAISC15\_J003531-434448&  0.69&18.23& 0.286&5& 9.92&10.71&$<$ 0.52&$<$ 0.56&$<$42.10&$<$42.14 \\ 
ELAISC15\_J003541-433302&  1.85&19.64& 0.716&3&11.57&11.48&$<$ 1.29&$<$ 1.60&$<$43.42&$<$43.52 \\ 
ELAISC15\_J003545-433216&  1.86&19.57& 0.399&4&10.76&10.59&$<$ 0.24&$<$ 0.31&$<$42.10&$<$42.21 \\ 
ELAISC15\_J003545-431833&  0.74&21.46&   ...&0&  ...&  ...&$<$ 0.36&$<$ 0.19&$<$43.62&$<$43.35 \\ 
ELAISC15\_J003548-430640&  0.80&19.49& 0.426&3&10.47&10.71&$<$ 0.89&$<$ 1.84&$<$42.74&$<$43.05 \\ 
ELAISC15\_J003550-430505&  1.00&20.13& 0.425&2&10.64&10.34&$<$ 1.48&$<$ 2.84&$<$42.95&$<$43.23 \\ 
ELAISC15\_J003603-433152&  4.59&21.11& 0.860&1&12.08&10.81&$<$ 0.38&$<$ 1.09&$<$43.07&$<$43.54 \\ 
ELAISC15\_J003612-434627&  1.18&24.27&   ...&0&  ...&  ...&$<$ 0.58&$<$ 1.11&$<$43.83&$<$44.12 \\ 
ELAISC15\_J003615-431327&  0.62&19.44& 0.330&5&10.04&10.40&$<$ 0.38&$<$ 1.24&$<$42.11&$<$42.62 \\ 
ELAISC15\_J003619-432608&  2.21&17.35& 0.106&3& 9.48&10.03&$<$ 0.66&$<$ 2.28&$<$41.27&$<$41.81 \\ 
ELAISC15\_J003620-430920&  0.76&  ...&   ...&0&  ...&  ...&$<$ 1.11&$<$ 1.84&$<$44.12&$<$44.34 \\ 
ELAISC15\_J003622-432826&  0.77&21.64& 0.863&2&11.32&11.02&$<$ 2.12&$<$ 2.15&$<$43.83&$<$43.84 \\ 
ELAISC15\_J003623-432702&  0.69&21.00& 0.590&3&10.76&10.60&$<$ 0.12&$<$ 1.21&$<$42.19&$<$43.20 \\ 
ELAISC15\_J003627-432426&  0.71&22.46&   ...&0&  ...&  ...&$<$ 1.05&$<$ 2.14&$<$44.09&$<$44.40 \\ 
ELAISC15\_J003640-433925&  0.98&21.59& 1.181&1&11.95&10.93&  1.47&  1.34& 44.20& 44.01         \\ 
ELAISC15\_J003647-433424&  0.74&  ...&   ...&0&  ...&  ...&$<$ 0.79&$<$ 2.10&$<$43.97&$<$44.39 \\ 
ELAISC15\_J003650-434519&  0.92&  ...&   ...&0&  ...&  ...&$<$ 0.37&$<$ 0.69&$<$43.64&$<$43.91 \\ 
ELAISC15\_J003656-434312&  1.06&17.90& 0.376&5&10.44&11.18&$<$ 1.03&$<$ 1.48&$<$42.67&$<$42.83 \\ 
ELAISC15\_J003648-432259&  1.57&24.35&   ...&0&  ...&  ...&$<$ 0.74&$<$ 1.86&$<$43.94&$<$44.34 \\ 
ELAISC15\_J003649-431018&  1.81&18.41& 0.194&4& 9.94&10.22&$<$ 0.97&$<$ 2.69&$<$42.00&$<$42.44 \\ 
\hline
\end{longtable}
\end{center}



\clearpage



\end{document}